\documentclass[aps,prl,twocolumn,showpacs,superscriptaddress,floatfix]{revtex4-1}

\usepackage{graphicx}
\usepackage{dcolumn}
\usepackage{bm}
\usepackage{amssymb}

\usepackage{microtype}
\begin{document}

\title{Incipient Ferromagnetism in Tb$_2$Ge$_2$O$_7$: Application of Chemical Pressure to the Enigmatic Spin Liquid, Tb$_2$Ti$_2$O$_7$}

\author{A.~M.~Hallas}
\affiliation{Department of Physics and Astronomy, McMaster University, Hamilton, ON, L8S 4M1, Canada}

\author{J.~G.~Cheng}
\affiliation{Beijing National Laboratory for Condensed Matter Physics, and Institute of Physics, Chinese Academy of Sciences, Beijing 100190, China}

\author{A.~M.~Arevalo-Lopez}
\affiliation{Centre for Science at Extreme Conditions and School of Chemistry, University of Edinburgh, King's Buildings, Mayfield Road, Edinburgh EH9 3JZ, United Kingdom}

\author{H.~J.~Silverstein}
\affiliation{Department of Chemistry, University of Manitoba, Winnipeg, MB, R3T 2N2, Canada}

\author{Y.~Su}
\affiliation{J\"{u}lich Centre for Neutron Science, Forschungszentrum J\"{u}lich GmbH, Outstation at MLZ, Lichtenbergstrasse 1, 85747 Garching, Germany}

\author{P.~M.~Sarte}
\affiliation{Department of Chemistry, University of Manitoba, Winnipeg, MB, R3T 2N2, Canada}

\author{H.~D.~Zhou}
\affiliation{Department of Physics and Astronomy, University of Tennessee, Knoxville, TN, 37996-1200, USA}
\affiliation{National High Magnetic Field Laboratory, Florida State University, Tallahassee, FL, 32306-4005, USA}

\author{E.~S.~Choi}
\affiliation{National High Magnetic Field Laboratory, Florida State University, Tallahassee, FL, 32306-4005, USA}

\author{J.~P.~Attfield}
\affiliation{Centre for Science at Extreme Conditions and School of Chemistry, University of Edinburgh, King's Buildings, Mayfield Road, Edinburgh EH9 3JZ, United Kingdom}

\author{G.~M.~Luke}
\affiliation{Department of Physics and Astronomy, McMaster University, Hamilton, ON, L8S 4M1, Canada}
\affiliation{Canadian Institute for Advanced Research, 180 Dundas St. W., Toronto, ON, M5G 1Z7, Canada}

\author{C.~R.~Wiebe}
\affiliation{Department of Physics and Astronomy, McMaster University, Hamilton, ON, L8S 4M1, Canada}
\affiliation{Department of Chemistry, University of Manitoba, Winnipeg, MB, R3T 2N2, Canada}
\affiliation{Department of Chemistry, University of Winnipeg, Winnipeg, MB, R3B 2E9 Canada}

\date{\today}

\begin{abstract}
The origin of the spin liquid state in Tb$_2$Ti$_2$O$_7$ has challenged experimentalists and theorists alike for nearly 20 years. To improve our understanding of the exotic magnetism in Tb$_2$Ti$_2$O$_7$, we have synthesized a chemical pressure analog, Tb$_2$Ge$_2$O$_7$. Germanium substitution results in a lattice contraction and enhanced exchange interactions. We have characterized the magnetic ground state of Tb$_2$Ge$_2$O$_7$ with specific heat, ac and dc magnetic susceptibility, and polarized neutron scattering measurements. Akin to Tb$_2$Ti$_2$O$_7$, there is no long-range order in Tb$_2$Ge$_2$O$_7$ down to 20~mK. The Weiss temperature of $-19.2(1)$~K, which is more negative than that of Tb$_2$Ti$_2$O$_7$, supports the picture of stronger antiferromagnetic exchange. Polarized neutron scattering of Tb$_2$Ge$_2$O$_7$ reveals that at 3.5~K liquid-like correlations dominate in this system. However, below 1~K, the liquid-like correlations give way to intense short-range ferromagnetic correlations with a length scale related to the Tb-Tb nearest neighbor distance. Despite stronger antiferromagnetic exchange, the ground state of Tb$_2$Ge$_2$O$_7$ has ferromagnetic character, in stark contrast to the pressure-induced antiferromagnetic order observed in Tb$_2$Ti$_2$O$_7$.
\end{abstract}


\maketitle
Geometrically frustrated pyrochlores, R$_2$M$_2$O$_7$, exhibit a diverse array of exotic magnetic behaviors \cite{GGG}. The ground states in these materials are dictated by a complex, and often delicate balance of exchange, dipolar, and crystal field energies. Tb$_2$Ti$_2$O$_7$ is one of the most remarkable of these frustrated pyrochlores; strong antiferromagnetic exchange and Ising-like spins led to predictions of an antiferromagnetic N\'eel state below $\sim$1~K for this material \cite{denHertog2000}. However, experimental studies revealed a lack of static order or spin freezing in Tb$_2$Ti$_2$O$_7$ down to 70~mK \cite{Gardner1999,Gardner2003}, and more recently 57~mK \cite{PhysRevB.86.020410}. Subsequently, enormous efforts have been undertaken to uncover the origin of the spin liquid state in Tb$_2$Ti$_2$O$_7$.

A further complication in Tb$_2$Ti$_2$O$_7$ is the coupling of magnetic and lattice degrees of freedom 
\cite{Aleksandrov1981,Ruff2007,Nakanishi2011,crystalfieldtbtio}. It has been suggested that hybridized magnetoelastic excitations may be responsible for the suppression of magnetic order in Tb$_2$Ti$_2$O$_7$ \cite{Fennell2014}. Another theoretical construct that attempts to account for the lack of static order in Tb$_2$Ti$_2$O$_7$ is a quantum spin ice state \cite{Molavian2007,PhysRevLett.105.047201,Fennell2012,Yin2013,Fritsch2013}. 
A third proposed scenario is that the non-Kramers doublet ground state of Tb$_2$Ti$_2$O$_7$ is split into two non-magnetic singlets through a symmetry reducing structural distortion \cite{PhysRevB.84.184409,PhysRevB.89.085115,PhysRevB.88.184428}. 

Other studies sought to uncover the origin of the spin liquid state in Tb$_2$Ti$_2$O$_7$ by focusing on mechanisms of its destruction, such as: external pressure \cite{Mirebeau2002}, magnetic fields \cite{Rule2006,PhysRevB.82.174406,PhysRevB.88.184428}, and a combination of the two \cite{Mirebeau2004}. Partial antiferromagnetic order is induced in Tb$_2$Ti$_2$O$_7$ with external hydrostatic pressures of 8.6~GPa, resulting in a 1\% compression of the lattice \cite{Mirebeau2002}. Another means of destroying the spin liquid state is chemical pressure: substitution of the non-magnetic titanium cation for an iso-electronic cation with a different ionic radius. Substitution of titanium in Tb$_2$Ti$_2$O$_7$ for the larger tin cation allowed exploration of negative chemical pressure \cite{Matsuhira2002}. In Tb$_2$Sn$_2$O$_7$, reduced antiferromagnetic exchange results in an ``ordered spin ice'' state at 0.87~K \cite{Mirebeau2005,Matsuhira2002}. In this two-in, two-out state, the spins are oriented 13.3$^{\circ}$ to the local $<$111$>$ axes. This ground state can be partially understood by a model of Heisenberg spins with finite ferromangetic exchange and $<$111$>$ anisotropy \cite{0295-5075-57-1-093}. However, this can only be reconciled with the apparent antiferromagnetic nearest neighbour exchange in Tb$_2$Sn$_2$O$_7$ if a tetragonal distortion is considered \cite{PhysRevB.85.054428}, for which there is currently no evidence. 

More recently, the study of chemical pressure has focused on substitution of Ti$^{4+}$ for the much smaller Ge$^{4+}$. Germanium substitution results in a lattice contraction and enhanced exchange interactions \cite{Zhou2012}. Study of the germanate pyrochlores has revealed that some frustrated ground states are stable against the application of chemical pressure while others are not. For example, the spin ice ground state is robust in the holmium pyrochlores, Ho$_2$\emph{B}$_2$O$_7$ (\emph{B} = Ge, Ti, Sn) \cite{Matsuhira2000,Hallas2012,Zhou2012}. Conversely, quantum fluctuations in the effective $S$~=~$1/2$ Yb$^{3+}$ cation are very sensitive to chemical pressure. Consequently, the ytterbium pyrochlores, Yb$_2$\emph{B}$_2$O$_7$ (\emph{B} = Ge, Ti, Sn), each have markedly different magnetic ground states \cite{Dun2014}.

To gain a better understanding of the exotic magnetism in Tb$_2$Ti$_2$O$_7$ we synthesized a positive chemical pressure analog, Tb$_2$Ge$_2$O$_7$. We characterized the magnetic ground state of Tb$_2$Ge$_2$O$_7$ with specific heat, magnetic susceptibility and polarized neutron scattering measurements. Akin to Tb$_2$Ti$_2$O$_7$, there is no long-range order in Tb$_2$Ge$_2$O$_7$ down to 20~mK. However, the liquid-like correlations in Tb$_2$Ge$_2$O$_7$ give way to intense short-range ferromagnetic interactions below 1~K. 

When reacted under ambient pressure, Tb$_2$Ge$_2$O$_7$ has a tetragonal pyrogermanate structure \cite{TbGeOPyrogermanate}. We prepared Tb$_2$Ge$_2$O$_7$ in the cubic pyrochlore phase using a high-temperature, high-pressure technique. Stoichiometric quantities of Tb$_2$O$_3$ and GeO$_2$ were reacted at 1000$^{\circ}$C and 8 GPa using a multi-anvil press. Batches of approximately 60~mg, prepared from a common precursor, were heated in rhenium capsules to produce a total of 320~mg of polycrystalline sample. Room temperature powder neutron diffraction measurements were made on the GEM diffractometer at the ISIS neutron facility. Rietveld fits to the diffraction pattern of Tb$_2$Ge$_2$O$_7$ with GSAS confirmed the Fd$\overline{3}$m pyrochlore phase and the absence of pyrogermanate impurities (Figure 1(a)). The room temperature lattice parameter was refined to a value of 9.9617(1) \AA. This corresponds to a reduction from Tb$_2$Ti$_2$O$_7$ and Tb$_2$Sn$_2$O$_7$ of $\sim$2\% and $\sim$5\% respectively (Table I). 

Off-stoichiometry and site-mixing are known to have a significant impact on the magnetic properties of Tb$_2$Ti$_2$O$_7$ \cite{Taniguchi2013,Fritsch20132} and other pyrochlores \cite{Ross2012}. These issues are most severe in the case of single crystals grown with the optical floating zone technique \cite{Ross2012}. A key advantage to the study of germanium pyrochlores is that the large size-mismatch between the rare earth and germanium cations should preclude the possibility of site-mixing. Rietveld refinement of the powder neutron diffraction pattern indicates ideal stoichiometry and the absence of site-mixing in Tb$_2$Ge$_2$O$_7$. The A and B sites of the lattice have an occupation of 1.00(5) by Tb$^{3+}$ and Ge$^{4+}$, respectively. 

\begin{figure}[htbp]
\includegraphics[width=3in]{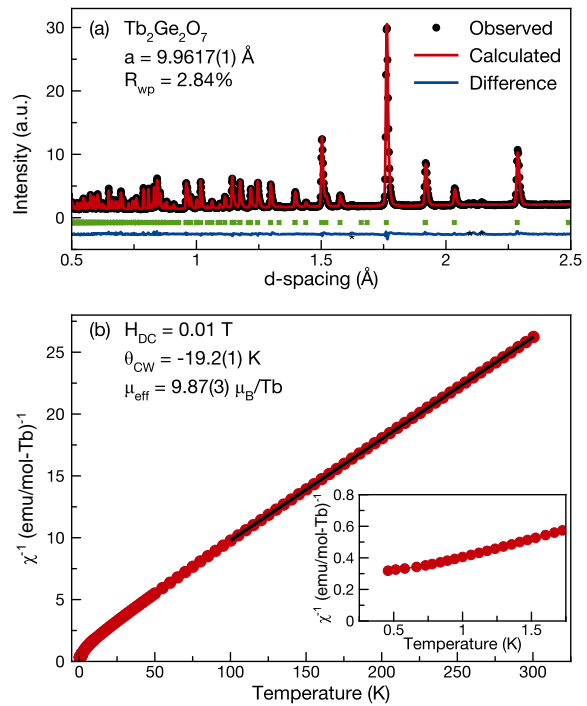}
\caption{(a) Neutron diffraction pattern of Tb$_2$Ge$_2$O$_7$ from the 90$^{\circ}$ detector bank confirming the pyrochlore structure (Fd$\overline{3}$m). Asterisks mark the reflections from vanadium and unreacted precursors. (b) dc susceptibility of Tb$_2$Ge$_2$O$_7$ where the black line is a Curie-Weiss law fit. Demagnetizing field effects were found to be negligible. Inset: Inverse susceptibility below 2~K.}
\label{Figure1}
\end{figure}

The inverse DC susceptibility of Tb$_2$Ge$_2$O$_7$ provides no evidence of long-range order down to 0.5~K (Figure 1(b)). A Curie-Weiss fit between 100~K and 300~K yields an antiferromagnetic Weiss temperature of $\theta_{\text{CW}}$~=~$-19.2(1)$~K. Fits over an identical temperature range for Tb$_2$Ti$_2$O$_7$ \cite{Gingras2000} and Tb$_2$Sn$_2$O$_7$ \cite{Matsuhira2002} give Weiss temperatures of $-17.5$~K and $-12.5$~K respectively. The more negative Weiss temperature for Tb$_2$Ge$_2$O$_7$ is indicative of stronger antiferromagnetic exchange that results from a reduced Tb-Tb distance. The susceptibility of Tb$_2$Ge$_2$O$_7$ begins to deviate from Curie-Weiss behavior below 70~K. This is similar to Tb$_2$Ti$_2$O$_7$ \cite{Gingras2000}, in which the deviation is attributed to the onset of developing short-range magnetic correlations \cite{Gardner1999}. 

\begin{table}[tbp]
\caption{Comparison of lattice and magnetic parameters in the Tb$_2$B$_2$O$_7$ (B = Sn \cite{Matsuhira2002,Mirebeau2005}, Ti \cite{Gingras2000}, Ge) pyrochlores.}
\begin{tabular}{lcccccccc}
\toprule
 && $a$ (\AA) && $\theta_{CW}$ (K) && $\mu_{\text{eff}}$ ($\mu_B$) && D$_{nn}$ (K) \\
\colrule
Tb$_2$Sn$_2$O$_7$ && 10.426 && $-12.5$ && 9.68 && 1.91 \\
Tb$_2$Ti$_2$O$_7$ && 10.149 && $-17.5(3)$ && 9.56 && 2.06 \\
Tb$_2$Ge$_2$O$_7$ && 9.9617(1) && $-19.2(1)$ && 9.87(3) && 2.19(1) \\
\botrule
\end{tabular}
\end{table}

The heat capacity of Tb$_2$Ge$_2$O$_7$ contains two low temperature anomalies centered at 5.5~K and 1.2~K (Figure 2). These features bear a significant qualitative resemblance to the low temperature heat capacity of Tb$_2$Ti$_2$O$_7$, which exhibits two peaks centered at 6~K and 1.5~K (inset of Figure 2). Gingras \emph{et al.} interpret the peak at 6~K in Tb$_2$Ti$_2$O$_7$ as a remnant of the first excited state doublet \cite{Gingras2000}. The anomaly at 1.5~K in Tb$_2$Ti$_2$O$_7$ is attributed to the build-up of short-range magnetic correlations. While the corresponding peak in Tb$_2$Ge$_2$O$_7$ is sharper, there is no evidence of the onset of long-range order at 1.2~K. We thus speculate that Tb$_2$Ge$_2$O$_7$ has a crystal field scheme which resembles that of Tb$_2$Ti$_2$O$_7$, resulting in similar heat capacity anomalies. In Tb$_2$Ti$_2$O$_7$, approximately $-6$~K of the Weiss temperature is due to crystal field effects \cite{Gingras2000}, 2~K are related to long-range dipolar contributions, and the remaining $-13.5$~K are due to magnetic exchange. The dipolar interaction, approximated as $D_{nn}~=~5/3(\mu_0/4\pi)\mu^2/r_{nn}^{3}$ in pyrochlores, does not significantly vary with lattice parameter (Table I). Given that the crystal fields and dipolar contributions are similar to those of Tb$_2$Ti$_2$O$_7$, we suggest that the increased magnitude of the Weiss temperature in Tb$_2$Ge$_2$O$_7$ results from enhanced antiferromagnetic exchange. However, a definite conclusion on this matter will require a detailed investigation of the crystal electric field of Tb$_2$Ge$_2$O$_7$.  

\begin{figure}[tbp]
\includegraphics[width=3in]{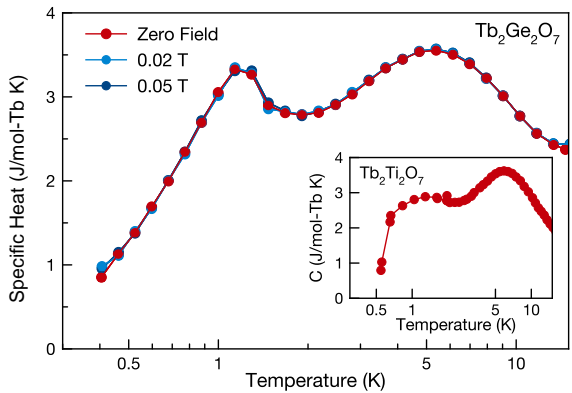}
\caption{The heat capacity of Tb$_2$Ge$_2$O$_7$ contains anomalies at 1.2~K and 5.5~K which strongly resemble the anomalies in the heat capacity of Tb$_2$Ti$_2$O$_7$ (inset, reproduced from \cite{Gingras2000}). The application of 0.02~T and 0.05~T fields does not alter these features.}
\end{figure}

We measured the ac susceptibility of Tb$_2$Ge$_2$O$_7$ down to 20~mK with frequencies ranging from 41~Hz to 511~Hz and in dc fields up to 0.05~T. In zero field, the real component of the susceptibility, $\chi^{\prime}$, 
contains no evidence of an ordering transition down to 20~mK in Tb$_2$Ge$_2$O$_7$ (Figure 3(a)). The imaginary component of the susceptibility, $\chi^{\prime\prime}$, is also free of anomalies and has an increasing magnitude with decreasing temperature (Figure 3(b)). The application of external dc fields as small as 0.01~T to Tb$_2$Ge$_2$O$_7$ induce a broad peak in $\chi^{\prime}$ (Figure 3(a)). As the external dc field is increased, the peak flattens and shifts to higher temperatures. At constant field strength, this feature is independent of frequency (Figure 3(c)). There is no corresponding peak or significant difference in $\chi^{\prime\prime}$. If this field-induced feature in Tb$_2$Ge$_2$O$_7$ had antiferromagnetic origins, an increasing field would suppress the peak to lower temperature. The field enhancement of this peak combined with its frequency independence suggest that it is ferromagnetic in character. This feature may be related to the formation of short-range spin correlations in Tb$_2$Ge$_2$O$_7$.

\begin{figure}[tbp]
\includegraphics[width=3.3in]{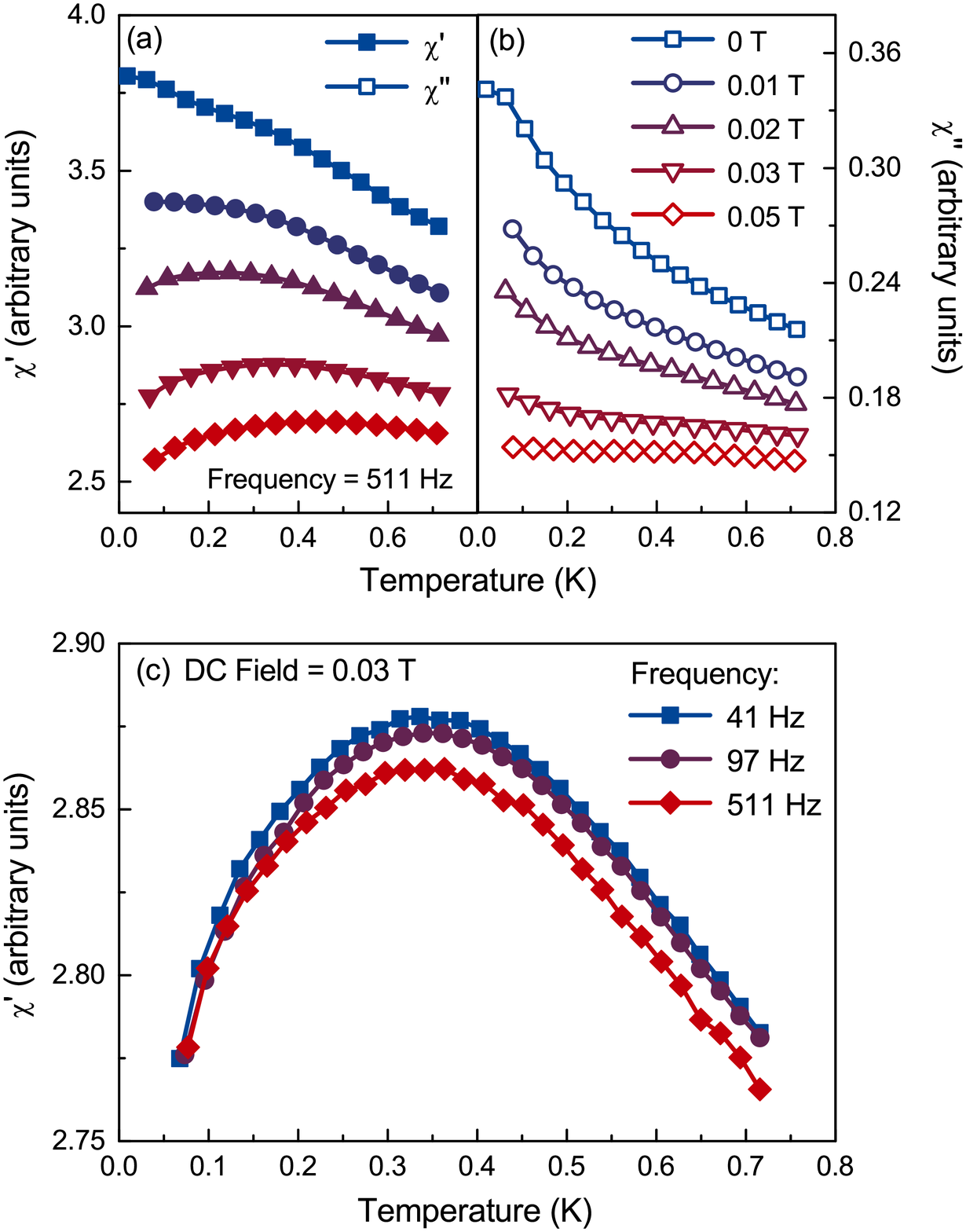}
\caption{The (a) real, $\chi^{\prime}$, and (b) imaginary, $\chi^{\prime\prime}$, components of the ac susceptibility of Tb$_2$Ge$_2$O$_7$. At zero field, there is no peak in the susceptibility down to 20~mK. (c) Small external dc fields induce a frequency independent peak that shifts to higher temperatures with increasing field.}
\label{Figure2}
\end{figure}

The related pyrochlores, Tb$_2$Ti$_2$O$_7$ and Tb$_2$Sn$_2$O$_7$, have been extensively characterized with ac susceptibility \cite{Ueland2006,Hamaguchi2004,Dahlberg2011,Yin2013}. In Tb$_2$Ti$_2$O$_7$, frequency dependent peaks at 350~mK and 140~mK are attributed to defect freezing \cite{Gardner2003} and a quantum spin ice state \cite{Yin2013} respectively. In Tb$_2$Sn$_2$O$_7$, the ordering transition at 850~mK is marked by a frequency independent feature in both the real and imaginary parts of the ac susceptibility \cite{Dahlberg2011}. With external magnetic field, this feature of Tb$_2$Sn$_2$O$_7$ is reduced in magnitude while shifting to higher temperatures. Although no peak is observed in the zero-field susceptibility of Tb$_2$Ge$_2$O$_7$, the behaviors otherwise more closely resemble Tb$_2$Sn$_2$O$_7$. It is possible that at the lowest temperatures Tb$_2$Ge$_2$O$_7$ is approaching an ordering transition that is not accessible experimentally. A further similarity between Tb$_2$Sn$_2$O$_7$ and Tb$_2$Ge$_2$O$_7$ is the increasing magnitude of the imaginary susceptibility below 1~K, which has been attributed to increasing ferromagnetic correlations in Tb$_2$Sn$_2$O$_7$ \cite{Dahlberg2011}.

We carried out polarized neutron scattering measurements on Tb$_2$Ge$_2$O$_7$ at the DNS spectrometer, which is operated by the Heinz Maier-Leibnitz Zentrum at Garching. Measurements were taken at 100~mK, 3.5~K, 25~K and 100~K using a dry-type dilution insert and a top-loading CCR with an incident wavelength of 4.2 \AA. {\it XYZ}-polarization analysis allows the magnetic scattering to be separated from the nuclear-coherent and spin-incoherent components. The scattering of Tb$_2$Ge$_2$O$_7$ at 25~K and 100~K is well-fit by the square of the magnetic form factor for Tb$^{3+}$ (Figure 4(a)). Thus, the scattering at these temperatures is mainly paramagnetic. However, the deviations from the magnetic form factor are more pronounced at 25~K compared to 100~K due to the development of short-range correlations.

At 3.5~K, the magnetic diffuse scattering in Tb$_2$Ge$_2$O$_7$ strongly deviates from the magnetic form factor. There is an upturn in the scattering at low-$Q$ and a hump in the scattering centered at 1.1~\AA$^{-1}$ (Figure 4(a)). These two features have competing origins. The upturn at low-$Q$ is related to short-range ferromagnetic correlations. The hump at 1.1~\AA$^{-1}$ is related to liquid-like correlations. A coexistence of short-range ferromagnetic and liquid-like correlations has also been observed in Tb$_2$Sn$_2$O$_7$ at 1.2~K, above its ordering temperature \cite{Mirebeau2005}. The fit to the data was achieved by combining a Lorentzian function and an antiferromagnetic nearest-neighbor spin correlation function. The Lorentzian function, $I(Q)~=~\frac{A}{\pi}\frac{\kappa}{\kappa^2 + Q^2}$, fits the short-range ferromagnetic correlations \cite{Mirebeau2008}. The liquid like scattering is modeled by $I(Q)$~$\approx$~$\frac{\sin(Qr_{ij})}{Qr_{ij}}$, where $r_{ij}$ is the distance between spins at sites \textit{i} and \textit{j} \cite{Gardner1999}. The value of $r_{ij}$ was refined to 3.65(6)~\AA, which agrees well with the Tb-Tb nearest neighbor distance in Tb$_2$Ge$_2$O$_7$ of 3.52~\AA. The antiferromagnetic contribution to the scattering, which is responsible for the maximum at $Q$~=~1.1~\AA$^{-1}$ and the minimum at $Q$~=~2.1~\AA$^{-1}$, is strongly reminiscent of the scattering in Tb$_2$Ti$_2$O$_7$ at 2.5~K \cite{Gardner1999}.

\begin{figure}[tbp]
\includegraphics[width=3in]{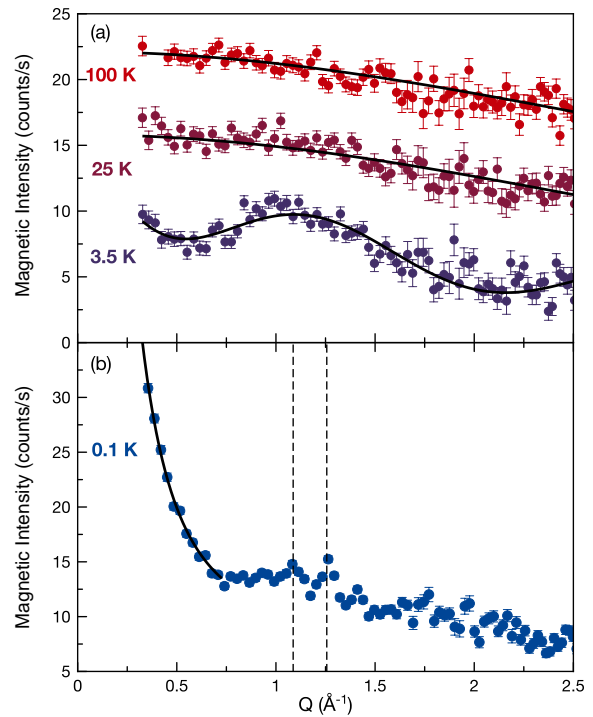}
\caption{Magnetic diffuse scattering of Tb$_2$Ge$_2$O$_7$ at (a) 100~K, 25~K, 3.5~K and (b) 100 mK. The scattering at 25~K and 100~K has been offset 6 and 12 counts/s respectively for clarity. The ferromagnetic Bragg peak positions (111) and (200) are indicated by the dashed lines. The black lines are the fits to data, as described in the text.}
\end{figure}

The $Q$-dependence of the magnetic diffuse scattering changes significantly between 100~mK and 3.5~K (Figure 4(b)). The decreased error bar size at 100~mK is due to significantly longer counting times. The spectral weight at 100~mK is increasing at low-$Q$ values, towards $Q$~=~0. The contribution from the liquid-like correlations, which were prominent at 3.5~K, are dwarfed by the low-$Q$ scattering. Thus, at 100~mK the magnetism in Tb$_2$Ge$_2$O$_7$ is dominated by short-range ferromagnetic correlations. This low-$Q$ scattering is fit to a Lorentzian function between 0.3~\AA$^{-1}$ and 0.7~\AA$^{-1}$. From this fit, a mean correlation length can be estimated by $\kappa^{-1}$ as 3.6(9) \AA, close to the Tb-Tb distance. Another feature at 100~mK is the presence of developing intensity at 1.08~\AA$^{-1}$ and 1.26~\AA$^{-1}$ (Figure~4(b)). These positions are the ferromagnetic Bragg peak positions (111) and (200). The (111) reflection is an allowed structural Bragg peak, but (200) is not. Examining the $\sim$60~hours of data collected at 100~mK reveals no change in intensity at these positions as a function of time. We thus consider two viable origins for this additional intensity: (i) An imperfect polarization analysis is giving rise to a residual signature from the nuclear channel or (ii) at 100~mK, Tb$_2$Ge$_2$O$_7$ is approaching ferromagnetic order that is static on the neutron time-scale. The absence of an ordering transition down to 20~mK in the ac susceptibility precludes the possibility of static order. While ac susceptibility can probe dynamics up to the kilohertz scale, neutrons are sensitive to dynamics on the order of terahertz.

It is worth noting that the 8.6~GPa of external pressure found to induce antiferromagnetic order in Tb$_2$Ti$_2$O$_7$ corresponds to a 1\% difference in lattice parameter \cite{Mirebeau2002}. Substitution of Ti$^{4+}$ by Ge$^{4+}$ represents a 2\% reduction in lattice parameter. Our results show that, despite enhanced antiferromagnetic exchange, Tb$_2$Ge$_2$O$_7$ does not order antiferromagnetically, nor is it even dominated by antiferromagnetic interactions. Thus, chemical pressure does not mimic the effects of external isotropic pressure in Tb$_2$Ti$_2$O$_7$. Chemical substitution of Tb$_2$Ti$_2$O$_7$ radically alters the magnetic ground state. This result further emphasizes the delicate balance of exchange, dipolar and crystal field interactions in these pyrochlores. Consideration of the similarities and differences between Tb$_2$Ti$_2$O$_7$ and Tb$_2$Ge$_2$O$_7$ should be useful for achieving a complete understanding of the origin of their collective paramagnetic states. To that end, a thorough study of Tb$_2$Ge$_2$O$_7$'s crystal field scheme will prove valuable. The local oxygen environment of terbium, which dictates the crystal electric field, is altered by germanium substitution. Our heat capacity measurements indicate qualitative similarities in the crystal fields of Tb$_2$Ge$_2$O$_7$ and Tb$_2$Ti$_2$O$_7$. However, it is possible that the significant differences in the magnetism of these two materials are related to subtle differences in the crystal field scheme and, consequently, the single ion anisotropy.

In conclusion, our results reveal a lack of long-range order in Tb$_2$Ge$_2$O$_7$ down to 20~mK. Magnetic diffuse neutron scattering measurements reveal that Tb$_2$Ge$_2$O$_7$ does not share a spin liquid ground state with Tb$_2$Ti$_2$O$_7$. Instead, Tb$_2$Ge$_2$O$_7$ is dominated by short-range ferromagnetic correlations with a length-scale characteristic of the Tb-Tb distance. A field induced peak with ferromagnetic character is observed in the ac susceptibility. Tb$_2$Ge$_2$O$_7$ represents an exciting new avenue to probe the exotic phase diagram of the terbium pyrochlores. Characterization of the crystal field scheme of Tb$_2$Ge$_2$O$_7$ and a $\mu$SR investigation of its dynamics will allow additional valuable comparisons to be drawn.

\begin{acknowledgments}
We are appreciative of helpful comments from M.~J.~P.~Gingras and B.~D.~Gaulin. We gratefully acknowledge the support received from staff at the Heinz Maier-Leibnitz Zentrum (MLZ), especially Kirill Nemkovskiy and Heinrich Kolb. We acknowledge support from EPSRC, STFC and the Royal Society. We thank Dr. W. Kockelmann (ISIS) for assistance with diffraction measurements. This work was supported by NSERC, the CRC program, and CFI. A portion of this work was performed at the NHMFL, which is supported by NSF Cooperative Agreement No. DMR-1157490, the State of Florida, and the U.S. DOE. J.G.C. acknowledges support from NSFC (Grant No. 11304371) and the Chinese Academy of Sciences (Grant No. Y2K5016X51).
\end{acknowledgments}

\bibliography{AHallas}

\end{document}